\documentstyle[aps,preprint]{revtex}

\begin{document}

\def\dx{\displaystyle}
\def\ang#1{\left \langle #1 \right \rangle}
\def\angg#1{\left \langle \left \langle  #1 \right \rangle  \right
\rangle} \def\const{\hbox{\rm const}}
\def\div{\hbox{\rm div}\, }
\def\curl{\hbox{\rm curl}\, }
\def\be{\begin{equation}}
\def\ee{\end{equation}}
\def\f{\frac}
\def\p{\partial}
\def\alf{\alpha}
\def\ba{{\bf a}}
\def\eps{\varepsilon}
\def\det{\hbox {\rm det}}
\def\d {\hbox {\rm d}}

\title {     Dynamics theory of deformable solids
	     with quasiparticle excitations in the presence
	     of electromagnetic fields
 }
\author{		  Dimitar I. Pushkarov}

\address{
		 Institute of Solid State Physics
      Bulgarian Academy of Sciences,
       Sofia 1784, Bulgaria,\\
       E-mail: dipushk@phys.acad.bg
 }
\maketitle

\begin{abstract}

	A full selfconsistent set of equations is deduced to describe
the kinetics
and dynamics of charged quasiparticles (electrons, holes etc.) with arbitrary
dispersion law in crystalline solids subjected to time-varying deformations.
The set proposed unifies the nonlinear elasticity theory equation, a kinetic
equation
for quasiparticle excitations and Maxwell's equations supplemented by the
constitute relations. The kinetic equation  used \cite{Push,AP}
 is valid for the
whole Brillouin zone. It is compatible with the requirement for
periodicity in
$k$-space and contains an essential new term compared to the traditional form
of the Boltzmann equation. The theory is exact in the frame of the quasiparticle
 approach
and can be applied to metals, semiconductors, as well as to other crystalline
solids including quantum crystals and low-dimensional lattice structures.
\end{abstract}


\section   { Introduction  }

There are two fundamental problems when dealing with quasiparticles
in crystalline structures. The first one is related to the fact that the
quasimomentum ${\bf k}$ and the dispersion law $\eps ({\bf k})$ of a
quasiparticle are well
defined only in an ideal periodic lattice. In such a lattice the dispersion law
as well as all other physical quantities are periodic functions in the
reciprocal space ($ k$-space). However, in any real system the crystal lattice
is deformed (e.g. by impurities, elastic deformations, external fields etc.).
The most complicated problems concern the quasiparticle dynamics in a
time-varying deformed crystal lattice. In an exact description all physical
characteristics of a quasiparticle have to be periodic functions of the
quasimomentum with periods which are functions of the coordinates and the
time. This leads to a dependence of the Brillouin zone boundaries  not
only on the deformation at a given instant, but also on the local lattice
velocity \cite{Push,AP}.
\par
As for the stationary (or quasistationary) case, this difficulty has
usually been passed over by
introducing a local dispersion law $\eps({\bf k}, u_{ik})$ and further
expansion in powers of the small deformation tensor components $u_{ik}$:
\begin{equation}
\eps({\bf k}, u_{ik}) = \eps_0({\bf k}) + \lambda_{ik}({\bf k})u_{ik}
\ee		     %1
where $\lambda_{ik}$ are the deformation potential components,
\begin{equation}
u_{ik}({\bf r}) = \frac {1}{2}\left( \frac{\partial u_i}{\partial x_k} +
\f{\p x_i}{\p x_k} \right),
\ee			   %2
and $\bf u$ is the deformation vector.
 Such an approach (known as the local lattice approach) is based on the
assumption that the
deformations are small and smooth functions of position and time and
possesses
all shortcomings of any linearized theory. In addition,
the kinetic equation
becomes incompatible with the requirement of the periodicity.
 To avoid
any misunderstandings, note that Eqs.(1) and (2)
are written in the co-moving frame.

	The second problem is related to the fact that the crystal lattice
plays the role of a privileged coordinate frame and no Galilean
transformations for quasiparticle characteristics exist.
 The lack of transformation laws for such quantities as energy,
Hamiltonian and quasimomentum means, in fact, that there is no consistent
quasiparticle mechanics. The most fundamental
quantity, the dispersion law, is known in a co-moving frame attached to
the lattice and this frame is even not inertial at time-varying
deformations. On the
other hand, all fundamental physical equations, such as conservation laws,
variational principles, kinetic eqs. etc. take place in the laboratory
frame ($L$-system). However  the concept of dispersion law does not exist in
$L$-system. Hence, in principle, even if  the mechanics equations for
a quasiparticle were known in $C$-system, they remain unknown in $L$-system.
\par
	These problems are so old as the quasiparticle approach itself is.
They
are well known, for example, in the theory of metals where many attempts have
been made to  derive a complete system of dynamic equations, consisting of
equations from the theory of elasticity, a kinetic equation for the electron
gas and Maxwell's
equations (cf. Refs. \cite{KontUFN,Kontbook}
 and the bibliography cited there).
\par
The attempts to deduce the equations of motion for a quasiparticle in
$L$-system can be divided into two groups corresponding to
 the twofold role of the
dispersion law. In the co-moving frame it coincides with both energy and
Hamiltonian. Therefore transformations typical of energy and Hamilton function
have been proved.
\par
    When considering $\eps({\bf k},u_{ik})$ as Hamiltonian a transformation
by a substitutional function of the form
\begin{equation}
\tilde \Phi({\bf r', p }, t) = ({\bf r' + u}({\bf r'}, t)){\bf p}
\ee
has been used and the following relations have been obtained
(see e.g. \cite{KontUFN,Kontbook}):
$$
{\bf k = p} + \frac{\p}{\p {\bf r'}}({\bf u p}),
\tilde H({\bf p,r}, t) =
\tilde \eps ({\bf k, r'},t) + {\bf \dot u p},
$$
where $\bf r = r' + u $ are the coordinates in $L$-system and $\, \bf p\,$
is supposed to be the corresponding quasimomentum.
However, $\tilde \Phi \, $ does not depend on the bare mass, $m$, of the
quasiparticle
and therefore does not take into account any inertial effects (e.g. the
 Stewart-Tolman
effect in metals, centrifugal forces in rotating bodies etc).
\par
If $\eps ({\bf k}, u_{ik})$ is considered as quasiparticle energy then
 the Galilean transformation applies as a consequence of
the requirement that energy density $\int \eps d^3 k$
(as a macroscopic quantity) has to obey Galilean transformations. This yields
\begin{equation}
\tilde {\cal E} = \eps_0 ({\bf k}) + \lambda_{ik}({\bf k})u_{ik} + m\dot{\bf u}
\frac{\p \eps_0}{\p {\bf k}}.
\ee
\par
It has been shown by many authors that such a transformation is incompatible
with the Boltzmann equation. That is why some artificial ways have been used
the most successful being that of Landau (cf the footnote in Ref.
\cite{KontUFN}). He
has suggested that (in order to take into account the noninertial properties of
 the lattice frame) one has first to add to the dispersion law (1)
the term $\dx -m\dot{\bf u}\frac{\p \eps}{\p {\bf k}}$, setting
\begin{equation}
\eps({\bf k}, u_{ik}) = \eps_0({\bf k}) + \lambda_{ik}({\bf k})u_{ik}
 -m {\dot {\bf u}}\f{\p \eps}{\p {\bf k}}.
\ee
 and then to apply the transformation by the substitutional function (3).
The result is:
\begin{equation}
\tilde H({\bf p, r}, t) = \eps_0({\bf k}) + \lambda_{ik}({\bf k})u_{ik}
+\left({\bf p} -m \f{\p \eps}{\p {\bf k}}\right)\dot {\bf u}.
\ee
\par
This procedure has been used in the most of
recent works. However it cannot be well-grounded due to its internal
inconsistency. In fact, if we consider the expression (5)
 as energy, and take a constant velocity ${\bf {\dot u}}$
then we come to the wrong conclusion that the energy in an inertial frame
could depend on the frame velocity. The same confusion follows
for the energy density which  must strictly obey
 the Galilean principle.
Note, that we may not consider (3) as a Hamiltonian, because the Hamilton
  function in $C$-system coincides with the dispersion law in accordance with
the Hamilton  equation
$ \dx
\dot {\bf r'} = {\bf v} = \frac{\p \eps}{\p {\bf k}}
$.
\par
These shortcomings have been removed in our previous works \cite{Push,AP,DSc}
 (see also
\cite{Singapore,Nauka})  where
transformations to replace the Galilean ones have been deduced and the
quasiparticle mechanics equations in Hamiltonian form have been presented.
This enabled us to derive  a Boltzmann-like kinetic equation valid in the whole
Brillouin zone. We have deduced
a selfconsistent set of equations for electrons in metals, taking
into
account some special features of the problem as the
 quasineutrality condition and the
neglection of the displacement current in Maxwell's eqs. The magnetic
permeability $\mu$ has also been taken constant ($\mu = 1$). These
approximations were good enough to develop
 the electron plasma hydrodynamics in crystalline
metals as well as to consider magnetohydrodynamic effects \cite{Klara}.
\par
The problem is more complicated in bad conductors and semiconductors, as well
as at higher frequences, when the displacement current cannot be neglected,
the quasineutrality condition does not hold and
$\epsilon$ and $\mu$ are functions of the
deformation. We give in this
work the solution of this general case.
The only assumption is
 that the constitute relations are linear, i.e.
$ D_i = {\hat \epsilon}_{ik} E_k$,
$ B_i = {\hat \mu}_{ik}H_k$.
\par

	In the present work we shall consider electrons
having in mind that the theory is valid for
arbitrary quasiparticles. We suppose for simplicity that the crystal considered
 is isotropic in its undeformed state. This means that $\epsilon$ and $\mu$ are
taken scalars, but their derivatives with respect to coordinates are matrices
and depend on the lattice symmetry. It is easy to generalize all the
results for the case where $\epsilon$ and $\mu$ are tensors of second rank. No
essentially new results should be obtained for this case, but the corresponding
relations are more cumbersome.
\par

	Finally, we would like to note that the problem considered is related
to some old questions about the electromagnetic forces acting to the body,
some specific features of electrodynamics in moving media,
the form of electromagnetic stress tensor in condensed matter,
the role of momentum and quasimomentum etc. Since we
 derive the full set of dynamics equations all the forces are taken
into account in a selfconsistent way.
\par
	The paper is organized in the following way. In Sec. 2. we introduce
new variables (discrete coordinates and ivariant quasimomentum),
write evolution equations for
lattice vectors in the real and reciprocal spaces and express in new notation
metrical tensors, deformation tensor etc.
In Sec. 3. we reproduce briefly some of our previous results on
the Hamilton eqs. and kinetic equation we need for this work.
In Sec. 4 we deduce the full set of equations which describes the behaviour of
quasiparticles in deformable solids in electromagnetic fields.

\section{ Notation }
\label{sec:notation}

Following our previous works \cite{Push,AP} (see also
\cite{Singapore,Nauka,DSc})
we consider a lattice with
primitive vectors ${\bf a}_\alpha,\,\, \alpha = 1,2,3$ and introduce discrete
coordinates $N^\alpha$ as the number of steps (each being equal to ${\bf
a}_\alpha$) in the lattice from the origin to a given point. In this notation
the differential coordinates $\,d {\bf r \,}$ which are considered as
physically
infinitesimal (i.e. large compared to the lattice constants but small compared
to any macroscopic distance of interest) can be written in the form:
$$
d{\bf r} = {\bf a}_\alpha d N^\alpha + {\bf {\dot u}} d t
$$
or
$$
d N^\alpha = {\bf a}^\alpha  d{\bf r} - {\bf a}^\alpha {\dot {\bf u}}dt
$$
where ${\bf a}^\alpha$ are the reciprocal lattice vectors which satisfy the
relations
\begin{equation}
{\bf a}_\alpha {\bf a}^\beta = \delta^\beta_\alpha, \qquad
a_{\alpha i} a_k^\alpha = \delta_{ik}.
\ee					  %7

   It is seen from the above equations that
\begin{equation}
{\bf a^\alpha} = \nabla N^\alpha, \qquad
{\bf a_\alpha} = \f{\p {\bf r}}{\p N^\alpha}, \qquad
{\dot {\bf u}} = - {\bf a}_\alpha {\dot N}^\alpha .
\ee			    %8
	The time-evolution equations for the vectors $\ba_\alf$ and $\ba^\alf$
can be deduced from plain geometrical considerations (Appendix 1) and written
as follows \cite{Push}:
\begin{equation}
\dot{\bf a}_{\alpha} + (\dot{\bf u}\nabla) {\bf a}_{\alpha} -
({\bf a}_{\alpha}\nabla)
\dot{\bf u} = 0
\end{equation}				    %9
\begin{equation}
\dot{\bf a}^{\alpha} + \nabla ({\bf a}^{\alpha}\dot{\bf u}) = 0.
\end{equation}				  %10

	 In the notation used the metrical tensors in the real and
reciprocal space are, respectively:
\begin{equation}
g_{\alf \beta} = \ba_\alf \ba_\beta, \qquad g^{\alf \beta} = \ba^\alf
\ba^\beta.
\ee		       %11

	Then the lattice cell volume equals $g^{1/2}$, where
$g = \det\, g_{\alf \beta} $.
\par
	The relations between the components of the metrical tensors and the
deformation tensor $\, u_{ik}\,$ follow from the expression for the interval:
\begin{equation}
ds^2 = g_{\alf \beta} dN^\alf dN^\beta .
\ee			 %12

The squared interval between the same points in the undeformed crystal is
$$
ds_0^2 = \,\stackrel{\circ}{g}_{\alf \beta} dN^\alf dN^\beta .
$$
	    Hence
\begin{equation}
ds^2 - ds_0^2 = 2w_{\alf \beta} dN^\alf dN^\beta
\ee			   %13
where
\begin{equation}
w_{\alf \beta} = \f{1}{2}(g_{\alf \beta} - \,\stackrel{\circ}{g}_{\alf \beta})
\ee			    %14
plays the role of the deformation tensor in our notation. The invariance of the
interval yields:
\begin{equation}
w_{\alf \beta}dN^\alf dN^\beta = w_{ik}dx^i dx^k .
\ee			     %15

    Taking into account relations (7) and (8) one obtains
\begin{equation}
w_{ik} = w_{\alf \beta} a_i^\alf a_k^\beta .
\ee			     %16
	In order to obtain the components $u_{ik}$  of the tensor of small
deformations (2) as well as to find the relation between the deformation vector
${\bf u = r - r}_0$ and the discrete coordinates $N^\alf$ introduced, let us
note that the quantities $\p u_i/\p x_k$ obviously coincide with the matrix
elements $\alpha_{ik}$ which describe the coordinate transformations
\begin{equation}
x_i = \,\stackrel{\circ}{x}_i + \, \alpha_{ik}\stackrel{\circ}{x}_k
\ee			    %17
and, consequently, the lattice vector transformations:
\begin{equation}
a_{\alf i} = \,\stackrel{\circ}{a}_{\alf i} +
\alpha_{ik}\,\stackrel{\circ}{a}_{\alf k}   .
\ee			    %18
	If the deformations are small, then,
\begin{equation}
\ba_\alf = \,\stackrel{\circ}{\ba}_\alf + \delta \ba_\alpha, \qquad
\ba^\beta = \,\stackrel{\circ}{\ba}{\!}^\beta + \delta \ba^\beta.
\ee			    %19

	Multiplying these two equations and taking into account relations (7)
and (17)
yield in a linear approximation with respect to $\delta {\bf a}^\beta$
\begin{equation}
\delta a_{\alf i} = -(\,\stackrel{\circ}{a}_{\beta i} \delta a^\beta_k)
\quad \stackrel{\circ}{a}_{\alf k} = \alpha_{ik}\stackrel{\circ}a_{\alpha k} .
\ee			%20
\par
	On the other hand the change of the discrete coordinates owing to the
deformation can be written in the form:
\begin{equation}
N^\alf = N^\alf_0 - w^\alf .
\ee			    %21
   where $w^\alpha$ is the deviation from the value $N^\alpha_0$ in the ideal
undeformed lattice.
Taking into account (8) one obtains
\begin{equation}
   \ba^\alf = \nabla N^\alf = \,\stackrel{\circ}{\ba}\!{}\!^\alf - \nabla
w^\alf \ee			   %22
	and hence,
\begin{equation}
\delta a^{\alf}_{i} = - \f{\p w^\alf}{\p x_i} .
\ee					%23

It follows from (20) and (23) that
\begin{equation}
\f {\p u_i}{\p x_k} = \alf_{ik} = \,\stackrel{\circ}{a}_{\alf i}
\f{\p w^\alf}{\p x_k} = \f{\p \stackrel{\circ}{a}_{\alf i} w^\alf}{\p x_k}
\ee					  %24
and therefore the deformation vector ${\bf u}$ and the deformation tensor
$u_{ik}$ in this linear case
are related to their discrete coordinate analogies, $w^\alf$ and
 $\dx \,\stackrel{\circ}{w}_{\alf \beta} = \f{1}{2}\left( \f{\p w_\alf}{\p
N^\beta} + \f {\p w_\beta}{\p N^\alf}\right)$, in the following way:
\begin{equation}
{\bf u} = \,\stackrel{\circ}{\ba}_{\alf} w^\alf, \qquad
u_{ik} = \,\stackrel{\circ}{w}_{\alf \beta} \,\stackrel{\circ}{a}\!{}\!^\alf_i
\,\stackrel{\circ}{a}\!{}\!^{\beta}_k
\ee			     %25
where $w_\alf = \,\stackrel{\circ}{g}_{\alf \beta} w^\beta$ are the covariant
components of ${\bf w}$.
\par
In the same way one can obtain the full deformation tensor components [5]:
\begin{equation}
w_{\alf \beta} = \f {1}{2} \left( \,\stackrel{\circ}{\ba}_{\alf} \f {\p{\bf u}}
{\p N^\beta} + \,\stackrel{\circ}{\ba}_{\beta} \f {\p {\bf u}} {\p N^\alf} +
\f {\p {\bf u}} {\p N^\alf} \f {\p {\bf u}} {\p N^\beta} \right)
\ee				%26

\section{ Dynamics and Kinetics of Quasiparticles}
\par
The starting point when deriving Hamilton equations for quasiparticles is that
in the co-moving frame ($C$-system) the dispersion law
$\eps ({\bf k},{g^{\alpha \beta}})$
coincides both with Hamiltonian and with energy. Hence, the equations of
motion in C-system can be written in the form
\begin{equation}
{\dot {\bf r}}'= \frac{\p \eps}{\p {\bf k}}, \qquad {\dot {\bf k}}=
-\frac{\p \eps}{\p{ \bf r}'}.
\ee		      %27

We have to determine Hamiltonian $H({\bf p,r},t)$ as a function of the
coordinates $\bf r$ and quasimomentum $\bf p$ in L-system in a way to have
canonical equations:
\be
{\dot {\bf r}}= \frac{\p H}{\p \bf p}, \qquad {\dot {\bf p}}=
-\frac{\p H}{\p {\bf r}}  .
\ee		      %28

	According to the general theory of Hamilton mechanics, the Hamiltonian
and the momentum can be obtained as derivatives of the action $S({\bf r},t)$
with respect to the time $t$ and the coordinates ${\bf r}$. Equations (27) show
that the variables $\bf k$ and $\bf r'$ are canonically conjugate. However,
if one
considers $\bf r'$ as a continuos variable, then the corresponding quantity
$\bf
k$ has to be considered as momentum. Quasimomentum has to be conjugate to some
discrete coordinate. Such coordinates, $\bf N$ (with components $N^\alpha$),
were
introduced in Sec.2. Hence, we are able to take the action $S({\bf N},t)$
as a function of these coordinates and
the time. Then we can determine a Hamiltonian $H(
\mbox {\boldmath $\kappa$},{\bf
N}, t)$ and a new quasimomentum  {\boldmath $\kappa$} as follows:
\begin{equation}
H(\mbox {\boldmath $\kappa$},{\bf N}, t) = -\left(\frac {\partial S}{\partial
t}\right)_{\bf N}  \qquad
\kappa_\alpha =
\left(\frac {\partial S}{\partial N^\alpha}\right)_t .
\ee	       %29
	  We call
{\boldmath $\kappa$} {\it the invariant
quasimomentum}, because all
physical quantities, written as functions of {\boldmath $\kappa$}, are
periodic
with a constant period $2 \pi$ (not $2\pi { \bf a}^\alpha$).
\par
	Let us consider the local dispersion law $\eps (\mbox {\boldmath
$\kappa$},
g^{\alpha \beta})$ as a function of the invariant quasimomentum and the
metrical tensor $g^{\alpha \beta}$. Since it coincides with the Hamiltonian,
one has
\begin{equation}
\eps(\mbox{\boldmath $\kappa$},g^{\alpha\beta}, t) = -\left(\frac {\partial
S}{\partial t}\right)_{\bf N}  \qquad
\kappa_\alpha =
\left(\frac {\partial S}{\partial N^\alpha}\right)_t.
\ee	       %30
	Now let us consider a real electron which is executing a quasiclassical
motion. Its wave function in the new variables has the form
\begin{equation}
\psi ({\bf N},t) \sim
exp\left\{\frac{i}{\hbar}S_0({\bf N},t)\right\}
\ee		  %31
where
$\dx S_0({\bf N},t)$ is the classical action. The transformation law for the
action follows from the transformation properties of the phase of the wave
function under Galilean transformations [9]:
\begin{equation}
S = S_0 + m{\bf \dot u r} - m{\dot u}^2/2
\ee		 %32
where  $m$ is the mass of a free particle.
\par
	The Hamiltonian and the quasimomentum in the Laboratory frame can now
be obtained as follows \cite{AP}:
\begin{equation}
{\bf p} = \left(\frac {\p S}{\p {\bf r}}\right)_t = {\bf a}^\alpha k_\alpha +
m{\bf \dot u}
\ee		     %33

\begin{equation}
H({\bf p,r},t) = -\left(\f{\p S}{\p t}\right)_{\bf r} = \eps + {\bf p \dot u} -
\frac{m{\dot u}^2}{2}
\ee		    %34
where $\eps = \eps({\bf a}_\alpha({\bf p}-m{\bf \dot u}), g^{\alpha \beta})$ is
a periodic function of $\bf p$ with periods $2 \pi \hbar {\bf a}^\alpha$
determined by the reciprocal lattice vectors corresponding to the deformed
local lattice. This is the reason to call $\bf p$ {\it the quasimomentum of the
quasiparticle in $L$-system}.
\par
It follows from (33) that
\begin{equation}
k_\alpha = {\bf a}_\alpha ({\bf p}- m{\bf \dot u}) = {\bf ka}_\alpha.
\ee		      %35
	Hence, the invariant quasimomentum components are equal to the scalar
product of the usual quasimomentum $\bf k$ in $C$-system and the primitive
vectors of the locally deformed lattice.
\par
The energy $\cal E$ of a quasiparticle in $L$-system obeys the Galilean law:
\begin{equation}
{\cal E} = \frac{m {\dot u}^2}{2} + m {\bf \dot u}\frac{\p \eps}{\p {\bf p}} +
\eps =	\frac{m {\dot u}^2}{2} + {\bf p}_0 {\bf \dot u} + \eps
\ee			 %36
where $\dx {\bf p}_0= m \frac{\p \eps}{\p {\bf p}}$ is the average momentum (the
mass flow) in  C-system. \par
Equations (33)-(36) replace the Galilean transformations, which are not valid
for quasiparticles because of the priviged of the crystal lattice frame.
\par
We are able now to write the Boltzmann equation:
\begin{equation}
\frac{\p f}{\p t} + \frac {\p f}{\p {\bf r}} \frac {\p H}{\p {\bf p}}
-\frac{\p f}{\p {\bf p}}
\left(\frac{\p H}{\p {\bf r}} - {\bf F}\right) = \hat{ I} f
\ee			   %37
where $\hat{I}$ is the collision operator. Note, that this  equation
becomes well
defined for quasiparticles only after obtaining the Hamiltonian
and the Hamilton equations.
\par
It has been shown in \cite{AP} that this equation is compatible with the
requirement of periodicity. This  can be seen also from its form if
$
f(\mbox{\boldmath $\kappa$}, {\bf r}, t)$ is taken as a function
of the invariant quasimomentum
{\boldmath $\kappa$} and the quantities
$\bf p$ and $H$ are substituted from (33) and (34). This yields:
\begin{equation}
\frac{\d f}{\d t} + {\bf a}_{\alpha}\frac{\partial\eps}{\partial
\kappa_{\alpha}}\left(\frac{\partial f}{\partial {\bf
r}}\right)_{\mbox {\boldmath $\kappa$}} -
{\bf a}_{\alpha}\frac{\partial f}{\partial
\kappa_{\alpha}}\left\{m\frac{\d \dot{\bf u}}{\d t} +
\left(\frac{\partial\eps}{\partial {\bf r}}\right)_{\mbox {\boldmath $\kappa$}}
  - m
\frac{\partial\eps}{\partial  \kappa_{\beta}}{\bf  a}_{\beta}\times
\curl \dot{\bf u} - {\bf F}\right\} = \hat{I}f
\end{equation}		   %38
where $\dx  \frac{\d}{\d t} = \frac{\partial }{\partial t} + (\dot{\bf
u}\nabla)$
and all quantities are differentiated with respect to the coordinates and the
time at constant {\boldmath $\kappa$}.

     The term $m \dx \frac{\d\dot{\bf u}}{\d t}$ takes into account
noninertial properties	of  the local frame.
 This  is
the  term which is responsible for the Stewart-Tolman effect
in  metals.
\par
     The term
\begin{equation}
m\frac{\partial\eps}{\partial \kappa_{\beta}}
{\bf a}_{\beta}\times  \curl \dot{\bf u}
\end{equation}		    %39
is of {\it essentially new kind} and cannot be obtained in  linear  theories.
It is proportional to the bare mass $\,m\,$ and, hence, is also
responsible for noninertial effects.
\par
	The Lorentz force in our notation has the form
\begin{equation}
{\bf  F}  =  -e{\bf  E}  -  \frac{e}{c}\,
\frac{\partial\eps}{\partial
\kappa_{\alpha}}{\bf a}_{\alpha}\times {\bf B} - \frac{e}{c}
\dot{\bf u} \times {\bf B}
\end{equation}		     %40
where {\bf E} and {\bf B} are the strengths of the electric field and  magnetic
induction, respectively. Substituting (40) into (38) yields the kinetic
equation for charged quasiparticles with a charge $-e$ ( $e>0$):
\begin{equation}
\frac{\d f}{\d t}  +  {\bf a}_{\alpha}\left(\frac{\partial    f}{\partial
{\bf r}}\right)_{\mbox {\boldmath $\kappa$}}
\frac{\partial\eps}{\partial
\kappa_{\alpha}} -
{\bf a}_\alpha
\frac{\partial f}{\partial \kappa_{\alpha}}\left\{ \left(
\frac{\partial\eps}{\partial{\bf r}}\right)_{\mbox {\boldmath $\kappa$}} +
e\tilde {\bf E}+ \frac{e}{c}\, \frac{\partial\eps}{\partial
\kappa_{\beta}}
 {\bf a}_{\beta}\times \tilde {\bf B} \right\} = \hat{I}f ,
\end{equation}		 %41
where
\begin{equation}
e \tilde{ \bf E} = e{\bf E} + \frac{e}{c}\dot{\bf u} \times {\bf B}
+ m\frac{\d\dot{\bf u}}{\d t}
, \qquad \tilde {\bf B} = {\bf B} - \frac{mc}{e}\,
\curl\dot{\bf u} .
\ee			 %42

\par
The first expression in (42) consists of two parts --- the electrical force
$\dx \,e \bf E'$ where $\dx{\bf E}' = {\bf E} + \frac{1}{c} {\bf \dot u \times
B}\,$ is the field in the co-moving frame, and the inertial force
$\dx\, m \frac
{\d {\bf \dot u}}{\d t}\,$.
The second relation in (42) can also be written in the form
$\dx \, \tilde {\bf B} = \curl ({\bf A} - \frac {c}{e}m \dot {\bf u})\, $
where $\, \bf A \,$ is the vector potential in an agreement with the well-known
rule of replacement of the particle momentum in a magnetic field
$\dx {\cal P} \rightarrow {\cal P} - \frac{e}{c}\bf A\,$ where now
${\cal P} = m \bf {\dot u} \,$.

 \par
	We shall use also another form of the kinetic equation.
The reason is that
further on we need to integrate some physical quantities over the Brillouin
zone, take integrals by parts as well as differentiate with respect to the
coordinates and the time. However, the Brillouin zone boundaries are functions
not only of the deformation at given instant, but also on the local lattice
velocity. The integration over the Brillouin zone does not commutate with the
differentiation with respect to the time and the coordinates.
As a result some fluxes appear
through the zone boundaries. This effect is important for nonequilibrium
systems, open Fermi-surfaces as well as for other cases
when the partition function or
its derivatives do not vanish on the zone boundaries.
This inconvenience can be
passed over by introducing a renormalized partition function
\begin{equation}
\varphi(\mbox {\boldmath $\kappa$}, {\bf r},t) = f/\sqrt{g} .
\end{equation}			  %43
\par
     The kinetic equation for $\varphi(\mbox{\boldmath $\kappa$},
 {\bf
r},t)$ has the form
\begin{eqnarray}
\dot{\varphi}  &  +  &	\div \left\{\left(\dot{\bf u} + {\bf
a}_{\alpha}
\frac{\partial\eps}{\partial \kappa_{\alpha}}\right)\varphi\right\}
-{\bf  a}_{\alpha} \frac{\partial}{\partial \kappa_{\alpha}} \left\{\varphi
\left(\nabla\eps + m \frac{\d \dot{\bf
u}}{\d t}\right )\right. \nonumber \\
& - &\left. m\varphi \frac{\partial\eps}{\partial  \kappa_{\beta}}
{\bf a}_{\beta}\times \curl\dot{\bf u} -  {\bf
F}\varphi\right\}  = \hat{I}\varphi .
\end{eqnarray}			     %44

     In this notation the differentiation with respect to $t$  and {\bf r} is
carried out at constant {\boldmath $\kappa$}
 and hence commutates with  $\int d^3
\kappa$. Results obtained by such a procedure can easily be rewritten
in the	previously adopted variables by the following substitutions
$$
{\bf a}_\alpha \frac{\partial}{\partial \kappa_\alpha}
\leftrightarrow
\frac{\partial}{\partial{\bf k}} \leftrightarrow
\frac{\partial}{\partial{\bf p}}
$$
\begin{eqnarray}
\langle f\ldots  \rangle
& \equiv &
\int\frac{d^3p}{(2\pi\hbar)^3}f({\bf p},
{\bf r},t)\ldots
 = \frac{1}{\sqrt g}  \int\frac{d^3
\kappa}{(2\pi\hbar)^3}f({\bf p},{\bf r},t)\ldots   \nonumber \\
&  =  &  \int \frac{d^3   \kappa}{(2\pi\hbar)^3}
\varphi(\mbox {\boldmath $\kappa$}, {\bf r}, t) \ldots
\equiv
\langle \langle \varphi\ldots
\rangle \rangle \, ,
\end{eqnarray}			 %45
\begin{equation}
\int  d^3 r\ldots  = \int d^3 N^{\alpha} g^{1/2}  \ldots
\end{equation}			 %46

\section{ Conservation Laws and Dynamics Equations}
\par
   In order to avoid cumbersome expressions we shall make our
consideration in three steps. First we write the conservation laws and
 consider the problem in a  general form without taking
into account the explicit form of the Maxwell's equations and electromagnetic
forces. Then we consider Maxwell's equations and constitute relations for
moving media and finally we combine the results and obtain the full
selfconsistent set of equations.
\par
    We start with the  conservation laws.
\par
	The continuity equation for quasiparticles is
\begin{equation}
m\dot{n} + \div {\bf j}_0 = 0
\end{equation}
where
\begin{equation}
 n =\langle\langle\varphi\rangle \rangle = \langle f\rangle ,
  \quad {\bf j}_0 =
m\left\langle \frac{\partial H}{\partial {\bf p}}f\right  \rangle =
m\left\langle \frac{\partial\eps}{\partial {\bf p}}f\right  \rangle
+ m n {\bf \dot u}.
\end{equation}
\par\noindent
   This equation follows directly from the kinetic equation \cite{Push}.
\par
     The total mass current is
\begin{equation}
 {\bf J}_0   = \rho\dot{\bf u} + {\bf j}'_0 ,
\end{equation}
where
\begin{equation}
 {\bf j}'_0 =
m\left\langle \frac{\partial\eps}{\partial {\bf p}}f\right  \rangle
\end{equation}
		%50
is the quasiparticle mass current with respect to the lattice and
$\rho = \rho_0
+ mn$ is the full mass density written as a sum of the lattice mass density
$\rho_0$ and the quasiparticle mass density.
\par
     The  quantities $\rho$  and ${\bf J}_0$  satisfy the mass continuity
equation
\begin{equation}
\dot{\rho} + \div{\bf J}_0 = 0 .
\end{equation}
		%51
The full momentum ${\bf J}$ is a sum  of $ {\bf J}_0$  and  the
field momentum ${\bf g}$:
\begin{equation}
{\bf J} = {\bf J}_0 + {\bf g}.
\end{equation}		       %52
     Note, that in this case {\it the full momentum does not coincide
with the mass flow!\/}

     Our aim is to determine momentum and energy fluxes $\Pi_{ik}$ and $\bf Q$
in such a  way	as  to	satisfy  the  continuity  equation (51),  the
momentum conservation law
\begin{equation}
\dot{J}_i + \nabla_k\Pi_{ik}   = 0,
 \end{equation} 	    %53
and the energy conservation law
\begin{equation}
\dot{E} + \div{\bf Q} = 0 .
\end{equation}		 %54
The energy in $L$-system is given by the expression
\begin{equation}
E = - \frac{1}{2}\rho_0\dot{\bf u}^2 + E_0(g^{\alpha\beta}) + \langle \langle
{\cal E} \varphi\rangle \rangle  + W ,
\end{equation}	      %55
where $E_0(g^{\alpha\beta})$ is the strain energy in $C$-system, and
$W $ is the field energy.

     The time derivative of the energy (55) is then
\begin{eqnarray}
\dot{E} & = & \rho\dot{\bf  u}\ddot{\bf  u}  +	\frac{1}{2}\dot{\rho}\dot{u}^2
%+ \sigma_{\alpha\beta}a_i^{\alpha}a_k^{\beta}
%\frac{\partial\dot{u}_i}{\partial x_k}  - \dot{\bfu}\nabla E_0
+ \frac{\partial}{\partial t}\angg{ \eps\varphi}
 + m\ddot{\bf  u} {\bf
a}_{\alpha}\angg{\varphi\frac{\partial\eps}{\partial  k_{\alpha}}}
 \nonumber\\ & + & m\dot{\bf u}\dot{\bf a}_{\alpha}
 \angg{\varphi\frac{\partial\eps}{\partial k_{\alpha}}}
  +  m\dot{\bf	u}{\bf a}_ {\alpha}
\angg{\varphi\frac{\partial\dot{\eps}}{\partial
k_{\alpha}} }  +m\dot{\bf  u}{\bf a}_ {\alpha}
\angg{\dot{\varphi}\frac{\partial\eps}{\partial
k_{\alpha}} }  + \dot{W} .
\end{eqnarray}	    %56
	The time derivative of the elastic energy $E_0(g^{\alpha\beta})$ can be
taken in the following way.
\begin{equation}
{\dot E}_0 = \frac{\p E_0}{\p g^{\alpha\beta}}{\dot g}^{\alpha\beta} =
-\sigma_{\alpha \beta} {\dot a}^\alpha_i a^\beta_i
\ee		       %57
where
\begin{equation}
\sigma_{\alpha \beta} = -2 \frac{\p E_0}{\p g^{\alpha\beta}} .
\ee			%58
   Substituting $\,{\dot a}^\alpha_i\,$ from the evolution equation (10)
in (57) yields
\begin{equation}
{\dot E}_0 = \sigma_{\alpha \beta} a^\beta_i \left(
\frac{\p a^\alpha_i}{\p x_k} {\dot u}_k +
a^\alpha_k \frac {\p {\dot u}_k} {\p x_i} \right) .
\ee			%59
On the other hand
\begin{equation}
\nabla_k E_0 =	\frac{\p E_0}{\p g^{\alpha\beta}} \nabla g^{\alpha\beta} =
-\sigma_{\alpha \beta}	a^\beta_i \frac {\partial a^\alpha_i}{\p x_k} .
\ee			%60

It follows from (59) and (60) that
\begin{equation}
{\dot E}_0 =
 \sigma_{\alpha\beta}a_i^{\alpha}a_k^{\beta}
\frac{\partial {\dot u}_i}{\partial x_k}  - \dot{\bf u}\nabla E_0 .
\ee		     %61
\par
The time derivative of the full momentum (52) gives
\begin{equation}
0 = -{\bf{\dot J}}+ {\dot \rho}{\dot {\bf u}} + \rho\ddot{\bf u} +
m\dot{\bf a}_ {\alpha} \angg{ \frac{\partial\eps}{\partial
k_{\alpha}}\varphi}   +m{\bf a}_{\alpha} \ang{
\frac{\partial\eps}{\partial k_{\alpha}}\dot{\varphi} } +
m{\bf a}_{\alpha} \angg{ \frac{\partial\dot{\eps}}{\partial
k_{\alpha}}\varphi }  + \dot{\bf g} .
\end{equation}		    %62
Multiplying (62) by $-\dot{\bf u}\,$ and adding the result to  the right
hand side of (56) yield
\begin{eqnarray}
\dot{E}  =  \dot{\bf u}\dot{\bf J} - \frac{1}{2}\dot{\rho}\dot{\bf u}^2 +
\sigma_{\alpha\beta}a_i^{\alpha}a_k^{\beta} \frac{\partial\dot{u}_i}{\partial
x_k}  - m\ddot{\bf  u} {\bf a}_{\alpha}
\angg{
\varphi\frac{\partial\eps}{\partial k_{\alpha}}}
  -  \dot{\bf  u}\nabla E_0
   +
   \dot{W}  -
\dot{\bf    u}\dot{\bf g} + \frac{\partial}{\partial t}
\langle \langle \eps\varphi\rangle \rangle .
\end{eqnarray}	       %63

The last term is considered in Appendix 2.  Substituting  the
time derivatives ${\dot{\bf J}}$ and ${\dot \rho}$ by means of (53) and (51)
one obtains after cumbersome calculations:
\begin{eqnarray}
\dot{E} & + & \nabla_k\left\{\frac{1}{2}\rho \dot{u}^2\dot{u}_k     +
\dot{u}_i\left(\Pi_{ik} - \rho\dot{u}_i\dot{u}_k  +  E_0\delta_{ik}  +
\langle \langle \eps f\rangle \rangle \delta_{ik}\right) \right.
\nonumber\\ & - &\left.\frac{1}{2}m\dot{u}^2
\ang{\frac{\partial\eps}{\partial p_k}f}  + \ang{\eps
\frac{\partial\epsilon}{\partial p_k}f} \right\}  \nonumber\\ & = &
\frac{\partial\dot{u}_i}{\partial  x_k}\left\{\Pi_{ik}		  -
\rho\dot{u}_i\dot{u}_k + \sigma_{i k}
- \langle \lambda_{i k}  f\rangle    +
E_0\delta_{ik}	 \right.\nonumber\\
&  -  &\left.  \dot{u}_i   j_{0i}   -\dot{u}_kj_{0i}
 \frac{}{}\right\} + \langle \eps\hat{I}f\rangle  +
\ang{{\bf F}\frac{\partial\eps}{\partial {\bf p}}f}  +
\dot{W} - \dot{\bf u}\dot{\bf g}
\end{eqnarray}		     %64
where
\begin{equation}
\lambda_{ik} = 2 \frac{\p \eps}{\p g^{\alpha\beta}}a^\alpha_i a^\beta_k
\qquad
\sigma_{ik} = -2 \frac{\p E_0}{\p g^{\alpha\beta}}a^\alpha_i a^\beta_k .
\ee			    %65
     The last three terms in (64) describe  the  change  of  the  field
energy, field momentum and the effect of external forces.  They  depend
on the concrete type of interaction.
\par
	If there are no external fields then the last three terms in (64)
should be omitted and the energy and momentum fluxes are:
\begin{equation}
Q_i   =   E\dot{u}_i + \ang{\eps\frac{\partial H}{\partial
p_i}f}
 - \frac{1}{2}\dot{u}^2 J_i + \Pi_{ik} \dot{u}_k
\end{equation}	       %66
\begin{eqnarray}
\Pi_{ik}  =
 - (\sigma_{ik} + E_0\delta_{ik})
 + \rho\dot{u}_i\dot{u}_k
%\nonumber \\
 +
 \langle\lambda_{ik}f\rangle
-m \left \langle f\frac{\partial \eps}{\partial p_i}\frac{\partial
\eps}{\partial p_k}\right  \rangle
 + m \left \langle f\frac{\partial H}{\partial p_i}\frac{\partial
H}{\partial p_k} \right \rangle .
\end{eqnarray}	       %67
\par
    The momentum flux tensor consists of two parts, $L_{ik}$ and $P_{ik}$,
which correspond to the contributions of the lattice and quasiparticles
respectively:
\begin{equation}
L_{ik} = -(\sigma_{ik} + E_0 \delta_{ik}) + \rho_0 {\dot u}_i {\dot u}_k
\ee			 %68
\begin{equation}
P_{ik} = \langle \lambda^0_{ik} f\rangle +
m \left \langle f \frac{\p H}{\p p_i} \frac{\p H}{\p p_k}\right\rangle
\ee			    %69
where
$$
 \langle \lambda^0_{ik} f\rangle =
 \langle \lambda_{ik} f\rangle
- m \left \langle f \frac{\p \eps}{\p p_i}
\frac{\p \eps}{\p p_k} \right \rangle
$$
is the quasiparticle momentum flux tensor in the system of centre of mass
while $ \langle \lambda_{ik} f\rangle $ corresponds to the co-moving frame.
It can be shown (Appendix 3) that the sume $\sigma_{ik} + E_0 \delta_{ik}$
corresponds (but coinsides in linear approximation only) to the stress tensor of
 the linear elasticity theory and turns into
pressure for isotropic media.

     Finally, the equation of the elasticity theory for an elastic
crystalline body with quasiparticle excitations in the absence of external
fields takes the form:
\begin{equation}
\frac{\partial}{\partial
t}(\rho\dot{u}_i) = - \frac{\partial\Pi_{ik}}{\partial	x_k}
 - \frac{\partial j_{0i}}{\partial t}	.
\end{equation}		       %70

The last term in the right-hand side describes the force which appears when
 varying the quasiparticle mass current with respect to the lattice.

Let us now consider the effect of the electromagnetic field.
The Maxwell's equations are:
\begin{eqnarray}
\curl{\bf E} &=& -\frac{1}{c}\,  \frac{\partial{\bf B}}{\partial t}\,
,\qquad \curl {\bf H} = \frac{4\pi}{c}{\bf j}_e + \frac{1}{c}\,
\frac{\partial{\bf D}}{\partial t} \, ,
 \\	    %71
\div{\bf D} &=& 4\pi q,  \qquad \qquad \div{\bf B}  =  0,
\end{eqnarray}			 %72
where
${\bf j}_e = {\bf j}'_e + q \dot {\bf u}$,
$q = q_0 -e n $ is the charge density ($q_0$ being
the lattice charge) and
\begin{equation}
\quad  {\bf j}'_e = -\frac{e}{m}{\bf j}'_0 ,\quad e > 0
\end{equation}	   %73
is the electron current density in the co-moving frame.
\par
We have now to take into account the field
terms in (64) containing the densities
of the field energy $W$, the field momentum $\bf g$ and the Lorentz force $\bf
F$.

	Before going on we would like to mention that the extraneous charges
$q_0$ have to be considered as a separate system. They are not accounted
in the Boltzmann equation as well as in the Lorentz force (40). Therefore,
 one needs some additional equations. In metals, this additional equation is
the quasineutrality condition \cite{AP}
 $q_0 = en$. This is a good approximation also
for semiconductors at low frequencies. The behaviour of the systems of
electrons and other  charges in the crystal  depends on the problem
considered. This is not the aim of our work.
The presence of extraneous charges makes the whole system open and
 conservation laws (47)-(55) do not present a full system. One has to
take into account both mechanical work and that of the induced forces.
 If the only current carriers are electrons, then one has to put
$q = 0$. However, in order to keep the general form of Maxwell's equations we
shall compensate the missing  contribution
\footnote{ Actually,
 if one is only interested in stress-tensor components then charges and
current can be omitted (c.f. Ref. \cite{LLE}, \S 34).
}
to the time derivative of the field energy by a term
$\dx \dot w =  -\dot{\bf u}
\left( q{\bf E} + \frac{1}{c} {\bf j_e \times B}\right)$.
\par
Therefore, the full field contribution is:
\begin{equation}
{\dot W}_f =
\ang{{\bf F}\frac{\partial\eps}{\partial {\bf p}}f}  +
\dot{W} - \dot{\bf u}\dot{\bf g} + \dot w .
\ee		      %74

	Maxwell's eqs. (71) and (72) are written in $L$-system.
They have to be supplemented
by  constitute relations. However, one has to keep in mind that these
relations have their known simple form only in the co-moving frame.
In that frame one has:
\begin{equation}
{\bf D}' = \epsilon {\bf E}', \qquad {\bf B }' = \mu {\bf H}',
\ee		%75
where
\begin{equation}
{\bf D}' = {\bf D} + \frac{1}{c}{\bf v \times H} ,\qquad
{\bf E }' = {\bf E } + \frac{1}{c}{\bf v \times B}
\ee		  %76
\begin{equation}
{\bf B}' = {\bf B} - \frac{1}{c} {\bf v} \times{\bf E} , \qquad
{\bf H }' = {\bf H } - \frac{1}{c} {\bf v} \times{\bf D}
\ee		   %77
and $\bf v = {\dot u}$ is introduced for convenience.
\par
It is easy to see that
\begin{equation}
 {\bf v D} = {\bf v D}' = \epsilon {\bf v E}' = \epsilon {\bf v E},
\qquad
 {\bf v B} = {\bf v B}' = \mu {\bf v H}' = \mu {\bf v H} .
\ee		    %78
\par
Relations (75)-(78)
 are exact althogh (76) and (77) coincide in letter to the field
transformations  with an accuracy to $v/c$.\footnote{ The exact field
transformations are
\begin{eqnarray}
{\bf \tilde E} &=&
{\bf\gamma E} +(1-\gamma){(\bf E.n)n} + \gamma \frac{\bf v}{c} \times {\bf B}
,\qquad %\nonumber \\
{\bf \tilde B} =
{\bf\gamma B} +(1-\gamma){(\bf B.n)n} - \gamma \frac{\bf v}{c} \times {\bf E}
\nonumber \\
{\bf \tilde D} &=&
{\bf\gamma D} +(1-\gamma){(\bf D.n)n} + \gamma \frac{\bf v}{c} \times {\bf H}
,\qquad %\nonumber \\
{\bf \tilde H} =
{\bf\gamma H} +(1-\gamma){(\bf H.n)n} - \gamma \frac{\bf v}{c} \times {\bf D}
\nonumber
\end{eqnarray}
where
$$
{\bf n} = \frac{{\bf v}}{v}, \qquad \gamma = \left( 1- \frac{ v^2}{c^2}
\right)^{-1/2}
$$
\par
These fields obviously satisfy relations (78).
Then from the constitute relations in the co-moving frame $\dx
{\bf \tilde D} = \epsilon {\bf \tilde E}, \,\,
 {\bf \tilde B} = \mu {\bf \tilde H}$
one obtains immediately relations (75)-(77) in the form:
\begin{eqnarray}
{\bf D} + \frac{1}{c}{\bf v \times H}
= \epsilon \left\{{\bf E } + \frac{1}{c}{\bf v \times B}\right\}
,\qquad
{\bf B} - \frac{1}{c} {\bf v} \times{\bf E}
= \mu \left\{ {\bf H } - \frac{1}{c} {\bf v} \times{\bf D}\right\}
\nonumber
\end{eqnarray}
}
 \par
     When taking the time derivative of the field energy $\dot W$ in (74)
one has to keep in mind that the permeabilities $\mu$ and $\epsilon$
in a nonstationary deformed media are
functions of space and time. For example, the variation of the
electrical part of the field energy in the lattice frame is:
$$
 \delta W'_E =\frac{1}{4 \pi}{\bf E'\delta D'} =
\frac{1}{4\pi}\left\{ {\bf E'^2 \delta\epsilon +
\epsilon E'\delta E'}\right\} = \frac {{\bf E'^2}}{8 \pi}\delta\epsilon +
\delta \frac{\epsilon {\bf E^2}}{8 \pi}.
$$
Therefore, the full variation of $W$ in time can be written
in the form:
\begin{eqnarray}
\dot W &=& \frac{1}{4 \pi} ({\bf E \dot D + H \dot B}) =\nonumber \\
&=&
\frac{1}{4 \pi}\left( {\bf E'} - \frac {{\bf v}}{c}\times{\bf B}\right)
\left( {\bf \dot D'} - \frac {{\bf v}}{c}\times{\bf \dot H}\right)
+
\frac{1}{4 \pi}\left( {\bf H'} + \frac {{\bf v}}{c}\times{\bf D}\right)
\left( {\bf \dot B'} - \frac {{\bf v}}{c}\times{\bf \dot E}\right)
= \nonumber \\ &=&
\frac{\partial}{\partial t} \frac {\epsilon E'^2 + \mu H'^2}{8 \pi} +
\frac{E'^2}{8 \pi}\dot\epsilon +
\frac{H'^2}{8 \pi}\dot\mu + {\bf v (\dot g + \dot G)} +
\nonumber \\ &+&
\frac{\partial}{\partial t}\frac{\epsilon({\bf v\times E'})^2
+ \mu({\bf v \times H'})^2}{8\pi c^2} +
\frac{({\bf v \times E'})^2}{8\pi c^2}\dot \epsilon
+\frac{({\bf v \times H'})^2}{8\pi c^2}\dot \mu + 0(v^3/c^3)
\end{eqnarray}
where
\begin{equation}
{\bf g} = \frac {{\bf E \times H}}{4 \pi c}, \qquad
{\bf G} = \frac {{\bf D \times B}}{4 \pi c} .
\end{equation}

From here on we shall restrict our consideration within an accuracy
to $v/c$ (neglecting terms $0(v^2/c^2)$. Then
\begin{equation}
\dot W - {\bf v \dot g} =
\frac{\partial}{\partial t} \frac {\epsilon E'^2 + \mu H'^2}{8 \pi} +
\frac{E'^2}{8 \pi}\dot\epsilon +
\frac{H'^2}{8 \pi}\dot\mu + {\bf v  \dot G} .
\end{equation}
\par
 The quantities $\epsilon$ and $\mu$ are functions of the metrical
tensor $g^{\alpha\beta}$. So their time derivatives $\dot \epsilon$
and $\dot \mu$
can be treated in the same way as the derivative of $E_0$ (c.f. (57)-(61)).
This yields:
\begin{equation}
\dot\epsilon =- \epsilon_{ik}\frac{\p v_i}{\p x_k} - {\bf v }\nabla\epsilon,
\qquad
\dot\mu =- \mu_{ik}\frac{\p v_i}{\p x_k} - {\bf v} \nabla\mu,
\ee			%95
where
\begin{equation}
\epsilon_{ik} = 2\frac{\p \epsilon}{\p g^{\alpha\beta}}a^\alpha_i a^\beta_k,
\qquad
\mu_{ik} = 2\frac{\p \mu}{\p g^{\alpha\beta}}a^\alpha_i a^\beta_k.
\ee		       %96
\par
   Substituting (82) and (83) in (81) and making use of the Poynting
theoreme  one obtains
\begin{eqnarray}
\dot W- {\bf v \dot g} &=& -\div{\bf S'} - {\bf j}'_e {\bf E}' -
\left( \frac{{\bf E}'^2}{8 \pi} \epsilon_{ik} +
\frac{{\bf H}'^2}{8\pi}\mu_{ik} \right) \frac{\p v_i}{\p x_k}\nonumber\\
&-& {\bf v} \left(\frac{{\bf E}'^2}{8 \pi}\nabla\epsilon + \frac{{\bf
H}'^2}{8\pi}  \nabla \mu\right)
+{\bf  v}{\bf \dot G} .
\end{eqnarray}		       %98
\par
      Neglecting terms of the order $0(v^2/c^2)$ in (79) means
that one can replace $\dot {\bf G}$ by $\dot {\bf G}'$.
The time derivative $\dot {\bf G'}$ can be transformed using Maxwell's
equations in the co-moving frame. This yields:
\begin{equation}
{\bf v} \dot{\bf G'} = -
{\bf v} \left(q {\bf E'} +
\frac{1}{c} {\bf j}'_e \times {\bf B'}\right)
+ v_i \nabla_k t'_{ik}
 + {\bf v} \left(\frac{{\bf E}'^2}{8 \pi}\nabla\epsilon + \frac{{\bf
H}'^2}{8\pi}  \nabla \mu\right)
\ee			     %99
where
\begin{equation}
t'_{ik} =
\frac{\epsilon}{4 \pi} \left( E'_i E'_k -
\frac {E'^2}{2} \delta_{ik}\right) +
\frac{\mu}{4 \pi} \left( H'_i H'_k -
\frac {H'^2}{2} \delta_{ik}\right)
\ee			      %100
is the Maxwell's stress tensor in the co-moving frame.
It is easy to show using eqs. (76)-(78) that
$$
{\bf v} \left(q {\bf E'} +
\frac{1}{c} {\bf j}'_e \times {\bf B'}\right)
=
{\bf v} \left(q {\bf E} +
\frac{1}{c} {\bf j}_e \times {\bf B}\right) =
{\bf j_e E - j'_e E'}.
$$
\par
The term which contains Lorentz force in (74) can be calculated
by means of (40):
\begin{eqnarray}
\ang{{\bf F}\frac{\partial\eps}{\partial {\bf p}}f} &=&
{\bf E}\ang{-e\frac{\partial\eps}{\partial {\bf p}}f}-
\frac{e}{c}{\bf B}\ang
{\frac{\partial\eps}{\partial {\bf p}}\times
\frac{\partial\eps}{\partial {\bf p}}
f} + \frac{1}{c} {\bf \dot u \times B}\ang{-e \frac{\partial\eps}
{\partial {\bf p}}f}
\nonumber \\ &=&
{\bf j}_e'\left( {\bf E} + \frac{\bf \dot u}{c}\times {\bf B}\right)
=
 {\bf j_e' E'}.
\end{eqnarray}				 %101
\par
The same term, but with a negative sign exists also in (84). Therefore,
the total work related to the Lorentz force (the mechanical one
and that of
 the electromotive forces) equals zero as it should be.
The terms related to the extraneous charges cancel for the same
reason.
      Finally,
\begin{equation}
{\dot W}_f = -\div{\bf S}' + \nabla_k v_i t'_{ik} -
T'_{ik} \frac{\p v_i}{\p x_k}
\ee			    %102
where
\begin{equation}
T'_{ik} =
\frac{1}{4 \pi} \left\{\epsilon E'_i E'_k +
\frac {E'^2}{2}(\epsilon_{ik} -
\epsilon  \delta_{ik})
+
 \mu H'_i H'_k +
\frac {H'^2}{2}( \mu_{ik} - \mu \delta_{ik}) \right\}
\ee			     %103
\par
  Hence, one has to add the  term
\begin{equation}
Q_i^f =  S_i' - v_k t'_{ik} =  S_i - v_iW
\ee			     %104
to the energy flux density in (66), as well as the term $ \, -T'_{ik}$ to
the momentum flux tensor $\Pi_{ik}$ in (67).
\par
     The elasticity theory equation takes then the form:
\begin{equation}
\frac{\partial}{\partial
t}(\rho\dot{u}_i) =
- \frac{\partial L_{ik}}{\partial  x_k}
- \frac{\partial P_{ik}}{\partial  x_k}
+ \frac{\partial T'_{ik}}{\partial  x_k}
+\frac{m}{e} \frac{\partial j'_{ei}}{\partial t}
-\frac{\partial g_{i}}{\partial t} \,  .
\ee		       %105
\par
    The  current and the electromagnetic
stress tensor $\hat T'$ are written
in the lattice frame.
Note, that the term
with the electrical current represents, in fact, the electron
mass-flow (the momentum, associated with the current). That part
of the electron mass current, which moves together with the latice,
is included in $\rho$ in the left-hand side of the equation.
The electromagnetic momentum flux tensor in $L$-system
has the form:
\begin{equation}
T_{ik} = \frac{1}{4 \pi}\left(\epsilon E_iE_k + \mu H_iH_k\right)
+(\epsilon_{ik} - \epsilon\delta_{ik})
\frac{E^2}{8 \pi} +(\mu_{ik} - \mu\delta_{ik})
\frac{H^2}{8 \pi} + v_iG_k + v_kG_i - v_ig_k -v_kg_i
\end{equation}
Obviously, for small velocities one can replace $T'_{ik}$  by the
corresponding tensor $T_{ik}$ in $L$-system.
\par
	If electrons in a good conductor (metal) are considered,
then
the quasineutrality condition holds and the displacement current as well
as the field momenta $\bf g$ and $\bf G$ have to be put zero. Atually,
the two terms in the right-hand side of the Ampere's law can be considered as
expansion with respect to the electric field frequency: $\dx {\bf j} +
\frac{1}{4\pi}\frac{\partial {\bf D}}{\partial t} \approx \sigma{\bf E} +
\frac{\omega\epsilon}{4\pi}\bf E$. In metals $\sigma \gg \omega\epsilon/(4\pi)$,
and the dissplacement current can be neglected.
As a result, one has
\begin{equation}
T^{metal}_{ik} =
 \frac{\mu}{4 \pi} H_iH_k
 +(\mu_{ik} - \mu\delta_{ik})
\frac{H^2}{8 \pi}
\end{equation}
This tensor  contains an additional
term $\dx \frac
{H'^2}{8\pi}\mu_{ik}$ compared to our previous works
 \cite{AP,Singapore,Nauka,Klara},
 in which
the magnetic permeability has been taken constant. As shown in
Appendix 3 in case of a
noncrystalline body (e.g. fluid)  the quantities $\mu_{ik}$ and
$\epsilon_{ik}$ have to be replaced by
$\dx \left(\rho\frac{\partial\mu}
{\partial\rho}\right)_T\delta_{ik}$ and
$\dx \left(\rho\frac{\partial\epsilon}
{\partial\rho}\right)_T\delta_{ik}$, correspondingly (comp. Ref.
\cite{LLE},
 \S 56).

\section { Conclusion}
\par
	In this work we have considered dynamics and kinetics of
 charged quasiparticles with arbitrary dispersion relations in
deformable crystalline structures.
We have chosen the most general case of time-varying deformations
when the quasimomentum is not a good quantum number, and energy
does not coinside to the Hamiltonian.
We have
 derived a full selfconsistent set of
equations which consists of the nonlinear elasticity
theory equation
(91) the kinetic equation (41) or (44) and Maxwell's equations
supplemented by constitute relations. The kinetic equation is valid
for the whole Brillouin zone and contains a new term
responsible for some inertial effects.
The elasticity theory equation
is derived from the conservation laws written in the most general
form. The only approximation is that the
electromagnetic field transformations are taken with an
accuracy to $0(v^2/c^2)$. Any higher accuracy for the solid-state
theory now would be pointless. In such a way the theory presented
is exact in the frame of the quasiparticle approach. It can be used
for any material (metal, semiconductor, quantum crystal, low-dimensional
structures etc.) with linear relations between electromagnetic
fields. It is easy to write the corresponding set of equations for
more than one type of quasiparticles (e.g. electrons and holes).

   In case of electrons in metals the results obtained fit well to
those of our previous works \cite{AP,Singapore}.

\acknowledgements

The author is grateful to Professor F.Bassani for helpful discussions.
Support from  National Science Fund (Bulgaria) and Consiglio Nazionale
delle Ricerche (Italy) is acknowledged.

\section* { Appendix 1}
\label{sec*:A1}
\par
	The evolution equation (9) for the primitive vectors $\dx \bf a_\alpha$
can be deduced from the following consideration (we shall omit the subscript
$\alpha$ for conveninence). The lattice vector $\bf a(r,t)$ at instant $t$ is
defined
by the two lattice sites $\bf r_1(t)$ and $\bf r_2(t)$: $ \bf a = r_2 - r_1$.
After a time interval $\delta t$ the lattice vector changes to
${\bf a'}({\bf r'},t+\delta t) = {\bf r'_2 - r'_1}$. The new positions of the
lattice sites are obviously
$ {\bf r'_1 = r_1 + v(r_1)}\delta t$, and
${\bf r'_2 =  r_2 + v(r_2)}\delta t$ where the velocity of the
lattice site $\bf r$ in the moment $t$ is denoted by $\bf v(r)= {\dot u}(r)$.
It follows from the last two equations that
$$
 {\bf a' - a = (v(r_2) - v(r_1))\delta t}.
$$
	Taking into account that
$\bf v(r_2) = v(r_1 + a) = v(r_1) +
(a\nabla)v$, and ${\bf a' = a(r + v}\delta t, t +\delta t) = {\bf a } +
{\bf {\dot a}} + {\bf (v \nabla) a}$
one obtains
$$
{\bf {\dot a}} + \bf (v \nabla) a - (a \nabla) v = 0. \eqno (A1.1)
$$
This equation coinsides with (9) in the Notation. It conserves automatically
the lattice vectors lines. In fact, the $\bf a$-vector line conservation
condition consists in collinearity of $\bf a$ and the left-hand side of
(A1.1),\cite{Kochin}  i.e. in
$$
\left[{\bf {\dot a}} + \bf (v \nabla) a - (a \nabla) v\right]\times {\bf a} =
0. \eqno (A1.2)
$$
Hence, eq. (A1.1) describes deformations which do not break or cross
crystalline lines with equal $\alpha$. This means that in a crystal lattice
free of dislocations the three functions
$N^{\alpha}({\bf r}, t)$ are single-valued, and equation (A1.1) desdribes
completely the evolution of the lattice configuration.

	The evolution equation for the reciprocal lattice vectors $\bf
a^{\alpha}$ can be obtained from (A1.1) and the relation
$${\bf a}^{\alpha} =
\frac{\partial N^{\alpha}}{\partial{\bf r}}.
\eqno(A1.3)
$$
 The later follows directly from the expression for the
physically infinitisimal differential coordinates at given instant
$
d{\bf r = a_{\alpha}}d N^{\alpha}
$
and relations (7). Multiplying (A1.1) by ${\bf a}^{\alpha}_k$  yields
$$
   a^{\alpha}_k {\dot a}_{\alpha i} + a^{\alpha}_k ({\bf \dot u} \nabla)
a_{\alpha i} - a^{\alpha}_k a_{\alpha s} \nabla_s {\dot u}_i = 0
\eqno (A1.4)
$$
Taking into account relations (7) one has
$$
  a_{\alpha i} \left( {\dot a}^{\alpha}_k + ({\bf \dot u}\nabla) a^{\alpha}_k
\right) + \nabla_k {\dot u}_i = 0
\eqno(A1.5)
$$
Multiplying this equation with $a^{\beta}_i$ yields
$$
\bf {\dot a}^{\alpha} + \nabla(a^{\alpha} {\dot u}) = 0
\eqno(A1.6)
$$
In obtaining (A1.6) we have taken into account that
$
 \nabla_i a^{\alpha}_k - \nabla_k a^{\alpha}_i = 0
$
as a consequence of (A1.3).

Substituting (A1.3) into (A1.6) gives
$$
\nabla \left( {\dot N}^{\alpha} + {\bf a^{\alpha}{\dot u}}\right) = 0
$$
and hence,
$$
{\bf \dot u} = -{\bf a}_{\alpha}{\dot N}^{\alpha}, \quad
{\dot N}^{\alpha} = -{\bf \dot u a^{\alpha}}.
\eqno(A1.7)
$$
It follows from (A1.3) and (A1.7) that
$$
d N^{\alpha} = \frac {\partial N^{\alpha}}{\partial {\bf r}}d {\bf r} +
\frac{\partial N^{\alpha}}{\partial t} d t = {\bf a^{\alpha}} d{\bf r} -
{\bf a^{\alpha} \dot u} dt
\eqno (A1.8)
$$
  This expression coincides with that given in Notation. It can be written also
in the form used in the text:
$$
d{\bf r} = {\bf a}_\alpha d N^\alpha + {\bf {\dot u}} d t
$$

 \section* { Appendix 2}
\label{sec*:A2}
\par
To obtain the time derivarive of the quasiparticle energy density
one can use the kinetic equation (44) for
$\varphi(\mbox {\boldmath $\kappa$},{\bf r},t)$.
Since the fluxes through the Brillouin zone boundaries in
this notation equal zero
it can be written in its integrated by parts form
$$
\dot \varphi + \div \left\{ \left( \dot{\bf u} + {\bf a}_\alpha
\frac{\partial\epsilon}{\partial \kappa_\alpha} \right) \varphi \right\}
+ {\bf a}_{\alpha} \left\{ \nabla \epsilon +
 m \frac{\d { \dot {\bf u}}}{\d t}
 -  m \frac{\partial\epsilon}{\partial	\kappa_{\beta}}
{\bf a}_{\beta}\times \curl\dot{\bf u} - {\bf F}\right\}\varphi
\frac{\partial}{\partial \kappa_\alpha} = \hat{I}\varphi .
\eqno(A2.1)
$$
\par
We have to transform the expresssion
$$
\frac{\partial}{\partial t} \langle \epsilon f \rangle	  =
\langle \langle \dot{\epsilon}\varphi\rangle \rangle	   +
\langle \langle \epsilon\dot{\varphi}\rangle \rangle.
\eqno(A2.2)
$$
As the derivative with respect to $t$ is taken at constant
 {\boldmath $\kappa$} one can
use the same procedure, as when obtaining (61). This yields
$$
\dot{\epsilon} = \left(\frac{\partial\epsilon}{\partial
t}\right)_{\mbox {\boldmath $\kappa$}}
=\frac{1}{2}\lambda_{\alpha\beta}\dot{g}^{\alpha\beta} = -
\lambda_{\alpha\beta} a_i^{\alpha}a_k^{\beta}
\frac{\partial\dot{u}_i}{\partial x_k} -
\dot{u}(\nabla\epsilon)_{\mbox {\boldmath $\kappa$}} \, ,
\eqno(A2.3)
$$	    +
where
$$
\lambda_{\alpha\beta} =
2\frac{\partial\epsilon}{\partial  g^{\alpha\beta}}  =
\lambda_{\beta\alpha}
\eqno(A2.4)
$$
\par
Multiplying (A2.3) by $\varphi$ and (A2.1) by $\epsilon$ one obtains after
substituting into (A2.2):
\begin{eqnarray*}
\frac{\partial}{\partial  t}  \langle \epsilon	f
\rangle
 & = &
\langle \langle \epsilon\hat{I}\varphi\rangle \rangle
- \div \left(\dot{\bf u}\langle \langle \epsilon\varphi\rangle \rangle
+ {\bf a}_{\alpha\beta} \angg{\epsilon
\frac{\partial\epsilon}{\partial k_{\alpha}}\varphi } \right) +  {\bf
a}_{\alpha}
\angg{
 {\bf F} \frac{\partial\epsilon}{\partial k_{\alpha}}\varphi }
 \nonumber\\ & - & m\ddot{\bf u}{\bf a}_{\alpha}
\angg{ \frac{\partial\epsilon}{\partial
k_{\alpha}}\varphi }
  -\frac{\partial u_i}{\partial x_k} \left(m\dot{u}_k a_{\alpha i}
\angg{
\frac{\partial\epsilon}{\partial   k_{\alpha}}\varphi } +
a_i^{\alpha}a_k^{\beta}
\langle \langle \lambda_{\alpha\beta}\varphi \rangle \rangle
\right)   \nonumber\\
& = &
\langle \epsilon\hat{I}f\rangle   -  \div\left(\dot{\bf  u}
\langle \epsilon  f\rangle   +	\ang{\epsilon
\frac{\partial\epsilon}{\partial {\bf p}}f}  \right) + \ang{
{\bf F}
\frac{\partial\epsilon}{\partial {\bf p}}f }  - m\ddot{\bf u}
\ang{\frac{\partial\epsilon}{\partial {\bf p}}f}  \nonumber\\
& - & \frac{\partial\dot{u}_i}{\partial x_k} \left(m\dot{u}_k
\ang{\frac{\partial\epsilon}{\partial p_i}f}  + a_i^{\alpha}
a_k^{\beta}\langle \lambda_{\alpha\beta}f\rangle \right)
\end{eqnarray*}
where the rules (45) are used to replace the double brakets by single ones.
This expression is used when obtaining (64).
\par
We would like here
to pointed out how easily this formula has been obtained.
For comparisson, the expression which corresponds to (A2.2) in variables
${\bf p, r},t$ has the form:
$$
\left( \frac{\partial \epsilon}{\partial t}\right)_{\bf p} =
-\lambda_{\alpha \beta}a^\alpha_i a^\beta_k \frac{\partial {\dot u_i}}
{\partial x_k} - m \ddot{\bf u}\frac{\partial \epsilon}{\partial {\bf p}}
-\dot{\bf u} (\nabla \epsilon)_{\bf p}	+
\left( p_i \frac{\partial \epsilon}{\partial p_k} - m \ddot u_k \right)
\frac{\partial \dot u_i}{\partial x_k}.
$$
\par
This expression is both cumbersom and nonperiodic which creates additional
difficulties.

\section* { Appendix 3}
\label{sec:A3}
\par
The lattice contribution $L_{ik}$ to the
momentum flux density tensor $\Pi_{ik}$ is
given by (68). The term in the brakets can be written in the form:
$$
\sigma_{ik} + E_0\delta_{ik} =
\left( \sigma_{\alpha\beta}
+ E_0 g_{\alpha \beta} \right)a^\alpha_i a^\beta_k
\eqno (A3.1)
$$
where
$E_0(g^{\alpha \beta})$ is the strain energy per unit volume.
\par
In an isotropic medium the dependence of $E_0$ on $g^{\alpha\beta}$ is
reduced to a dependence on $g = \det g_{\alpha\beta}$:
$$
\sigma_{\alpha\beta} = -2\frac{\partial E_0}{\partial g}
\frac{\p g}{\p g^{\alpha\beta}}.
$$
 \par
By the well known formula
$$
d g = - g g_{\alpha\beta}dg^{\alpha\beta}
$$
one obtains easily
$$
\sigma_{\alpha\beta} =
2g \frac{\p E_0}{\p g}a_{\alpha l}a_{\beta l}
$$
and hence
$$
\sigma_{ik} =
2g \frac{\p E_0}{\p g}a_{\alpha l}a_{\beta l}
a^{\alpha }_i a^\beta_k =
2g \frac{\p E_0}{\p g}\delta_{ik}
\eqno(A3.2)
$$
\par
Taking into account that $g$ is the squared volume of a unit cell
($ g = V^2$) and that $V = 1/\rho$ one obtains
$$
\sigma_{ik} = -\rho \frac{\p E_0}{\p \rho}\delta_{ik}
\eqno(A3.3)
$$
\par
Hence, it is seen, that in an isotropic case
$$
2 a_{i}^\alpha a_k^{\beta}\frac{\partial}{\partial g^{\alpha \beta}}
\quad \rightarrow \quad \delta_{ik} \rho\frac{\partial}{\partial \rho}.
\eqno(A3.4) $$
\par
Let now $\tilde E_0$ and $\, s\,$ be the internal energy and entropy
per unit mass $\,(\tilde E_0 = E_0 V)\,$.
Making use of the thermodynamic relation
$$
d \tilde E_0 = T d s - P d V + \mu d N
\eqno(A3.5)
$$
one can define the pressure $\,P\,$ as
$$
P = -\left(\frac{\p \tilde {E_0}}{\p V}\right)_{s,N} =
 -\left(\frac{\p E_0 V}{\p V}\right)_{s,N} =
 -E_0 + \rho\left(\frac{\p E_0}{\p \rho }\right)_{s,N}
\eqno(A3.6)
$$
\par
It follows from (A3.3) and (A3.6) that in isotropic medium
$$
\sigma_{ik} + E_0 \delta_{ik} = -P\delta_{ik}
\eqno(A3.7)
$$
\par
It is supposed in our consideration that the only contribution to the
entropy is due to quasiparticles and this contribution comes from the
kinetic equation. Therefore, the derivatives of $\, E_0 \,$ with respect
to the metrical tensor components $g^{\alpha\beta}$ are assumed as taken
 at constant entropy.
\par
An alternative approach can be based on the free energy thermodynamic
potential per unit volume $\, F(T,P,N)\,$. In that case one obtains
$$
-P = F - \rho\left(\frac{\p F}{\p \rho }\right)_{T}
\eqno(A3.8)
$$


\begin{references}

\bibitem{Push} D.I. Pushkarov,
  J.Phys.C- Solid State Phys. {\bf 19} 6873 (1986);
 Preprint P17-85-224, Joint Inst. Nucl. Res., Dubna, USSR (1985).
\bibitem{AP}
A. F. Andreev  and  D. I. Pushkarov,   Zh. Eksp. Teor. Fiz.
{\bf 89} 1883 (1985)  (in Russian),
(Sov. Phys.JETP {\bf 62} 1087 (1985)).
\bibitem{KontUFN}
V. M. Kontorovich,    Usp. Fiz. Nauk {\bf 142}(2) (1984) 265.
(in Russian)
\bibitem{Kontbook}
V. M. Kontorovich,  In {\it Elektroni  Provodimosti},  eds.
M. I. Kaganov and V. S. Edelman, Moscow, Nauka, 1985.	(in Russian.)
\bibitem{Singapore}
 D. I. Pushkarov, {\it Quasiparticle Theory of	Defects  in
Solids},
World Scientific, Singapore-New Jersey-London-Hong Kong, 1991.
\bibitem{Nauka}
 D. I. Pushkarov,  {\it Defektony v Kristallakh \ldots}
( Defectons  in  Crystals.
Quasiparticle Approach to Defects in Solids), Moscow, Nauka 1993. (in
Russian.)
\bibitem{DSc}
 D. I. Pushkarov, {\it Quasiparticle  Approach	in  Quantum  Theory
of Solids}, DSc Thesis, Dubna, USSR, 1986. (in Russian.)
\bibitem{Klara}
D.I.Pushkarov, R.D.Atanasov and K.D.Ivanova,
{\it Phys.Rev.}{\bf B 46} (12) 7374-7378 (1992)
\bibitem{LLQ}
L. D. Landau and E. M. Lifshitz,  {\it Quantum Mechanics},
Nauka, Moscow, (problem in \S17), 1974.  (in Russian.)
\bibitem{LLE}
L. D. Landau and E. M.	Lifshitz,  {\it Electrodynamics of
Continuous Media}, Nauka, Moscow \S65, 1978.  (in Russian.)
English translation: Pergamon Press (1963)
\bibitem{Kochin}
N. E. Kochin, {\it Vector Calculus and the Rudiments of Tensor Calculus},
Nauka, Moscow 1965 (in Russian).

\end{references}
\end{document}